\documentclass{llncs}

\usepackage[english]{babel}
\usepackage{url}
\usepackage{enumerate}
\usepackage{wrapfig}
\usepackage{bbm}
\usepackage{flushend}
\usepackage[capitalise]{cleveref}
\usepackage{color}
\usepackage{float}
\usepackage[caption=false]{subfig}
\usepackage{graphicx}
\usepackage{amsfonts}
\usepackage{amssymb}
\usepackage{mathtools}
\usepackage{multirow}
\usepackage{longtable}

\usepackage{calc}
\renewcommand\thetable{\arabic{table}}

\newtheorem{thm}{Theorem}

\newtheorem{lem}[thm]{Lemma}

\newcommand{\var}[1]{\mbox{\textit{var}}\left(#1\right)}
\newcommand{\E}[1]{\mathbb{E}\left[#1\right]}
\newcommand{\prob}[1]{\mathbb{P}\left(#1\right)}
\normalsize

\begin{document}

\title{Modelling of trends in Twitter using retweet graph dynamics}
\titlerunning{Modelling of trends in Twitter} 

\author{Marijn ten Thij\inst{1} \and Tanneke Ouboter\inst{2} \and
Dani\"{e}l Worm\inst{2} \and Nelly Litvak \inst{3}\thanks{The work of Nelly Litvak is partially supported the EU-FET Open grant NADINE (288956)} \and Hans van den Berg\inst{2,3} \and Sandjai Bhulai\inst{1}}

\authorrunning{Marijn ten Thij et al.} 

\tocauthor{Marijn ten Thij, Tanneke Ouboter, Dani\"{e}l Worm, Nelly Litvak, Hans van den Berg and Sandjai Bhulai}

\institute{VU University Amsterdam, Faculty of Sciences, the Netherlands,\\
\email{\{m.c.ten.thij,s.bhulai\}@vu.nl},
\and
TNO, Delft, the Netherlands,\\
\email{\{tanneke.ouboter,daniel.worm,j.l.vandenberg\}@tno.nl}
\and
University of Twente, Faculty of Electrical Engineering, Mathematics and Computer Science, the Netherlands\\
\email{n.litvak@utwente.nl}
}
\maketitle
\begin{abstract}

In this paper we model user behaviour in \emph{Twitter} to capture the emergence of	 trending topics. For this purpose, we first extensively analyse tweet datasets of several different events. In particular, for these datasets, we construct and investigate the retweet graphs. We find that the retweet graph for a trending topic has a relatively dense largest connected component (LCC). Next, based on the insights obtained from the analyses of the datasets, we design a mathematical model that describes the evolution of a retweet graph by three main parameters. We then quantify, analytically and by simulation, the influence of the model parameters on the basic characteristics of the retweet graph, such as the density of edges and the size and density of the LCC. Finally, we put the model in practice, estimate its parameters and compare the resulting behavior of the model to our datasets. 

\end{abstract}


\section{Introduction}\label{sec:intro}
Nowadays, social media play an important role in our society. The topics people discuss on-line are an image of what interests the community. Such trends may have various origins and consequences: from reaction to real-world events and naturally arising discussions to the trends manipulated e.g. by companies and organisations~\cite{ratkiewicz2011truthy}. Trending topics on \emph{Twitter} are `ongoing' topics that become suddenly extremely popular\footnote{\url{https://support.twitter.com/articles/101125-about-trending-topics}}. In our study, we want to reveal differences in the retweet graph structure for different trends and model how these differences arise.

In \emph{Twitter}\footnote{\url{www.twitter.com}} users can post messages that consist of a maximum of 140 characters. These messages are called tweets. One can ``follow'' a user in \emph{Twitter}, which places their messages in the message display, called the timeline. Social ties are directed in \emph{Twitter}, thus if user A follows user B, it does not imply that B follows A. People that ``follow'' a user are called ``friends'' of this user. We refer to the network of social ties in \emph{Twitter} as the friend-follower network. Further, one can forward a tweet of a user, which is called a retweet.

There have been many studies on detecting different types of trends, for instance detecting emergencies \cite{klein2012detection}, earthquakes \cite{sakaki2010earthquake}, diseases \cite{paul2011you} 
or important events in sports \cite{lanagan2011using}. In many current studies into trend behaviour, the focus is mainly on content of the messages that are part of the trend, see e.g.~\cite{lehmann2012dynamical}. Our work focuses instead on the underlying networks describing the social ties between users of \emph{Twitter}. Specifically, we consider a graph of users, where an edge means that one of the users has retweeted a message of a different user.

In this study we use several datasets of tweets on multiple topics. First we analyse the datasets, described in Section~\ref{sec:data}, by constructing the retweet graphs and obtaining their properties as discussed in Section~\ref{sec:retweet graphs}. Next, we design a mathematical model, presented in Section~\ref{sec:model}, that  describes the growth of the retweet graph. The model involves two attachment mechanisms. The first mechanism is the preferential attachment mechanism that causes more popular messages to be retweeted with a higher probability. The second mechanism is the superstar mechanism which ensures that a user that starts a new discussion receives a finite fraction of all retweets in that discussion~\cite{bhamidi2012twitter}. We quantify, analytically and with simulations, the influence of the model parameters on its basic characteristics, such as the density of edges, the size and the density of the largest connected component. In Section~\ref{sec:practice} we put the model in practice, estimate its parameters and compare it to our datasets. We find that what our model captures, is promising for describing the retweet graphs of trending topics. We close with conclusions and discussion in Section~\ref{sec:conclusion}.

\section{Related work}\label{sec:related}
The amount of literature regarding trend detection in \emph{Twitter} is vast. The overview we provide here is by no means complete. Many studies have been performed to determine basic properties of the so-called ``Twitterverse''. Kwak et al. \cite{kwak2010twitter} analysed the follower distribution and found a non-power-law distribution with a short effective diameter and a low reciprocity. Furthermore they found that ranking by the number of followers and PageRank both induce similar rankings. They also report that \emph{Twitter} is mainly used for News (85\% of the content). Huberman et al. \cite{huberman2008social} found that the network of interactions within \emph{Twitter} is not equal to the follower network, it is a lot smaller.

An important part of trending behaviour in social media is the way these trends progress through the network. Many studies have been performed on \emph{Twitter} data. For instance, 
\cite{bhattacharya2012sharing} studies the diffusion of news items in \emph{Twitter} for several well-known news media and finds that these cascades follow a star-like structure. Also, \cite{zhou2010information} investigates the diffusion of information on \emph{Twitter} using tweets on the Iranian election in 2009, and finds that cascades tend to be wide, not too deep and follow a power law-distribution in their size. 

Bhamidi et al. \cite{bhamidi2012twitter} proposed and validated on the data a so-called superstar random graph model for a giant component of a retweet graph. Their model is based on the well-known preferential attachment idea, where users with many retweets have a higher chance to be retweeted~\cite{barabasi1999emergence}, however, there is also a superstar node that receives a new retweet at each step with a positive probability. We build on this idea to develop our model for the progression of a trend through the \emph{Twitter} network.

Another perspective on the diffusion of information in social media is obtained through analysing content of messages. For example, \cite{sadikov2009information} finds that on \emph{Twitter}, tags tend to travel to more distant parts of the network and URLs travel shorter distances. 
Romero et al. \cite{romero2011differences} analyse the spread mechanics of content through hashtag use and derive probabilities that users adopt a hashtag.

Classification of trends on \emph{Twitter} has attracted considerable attention in the literature. Zubiaga et al. \cite{zubiaga2011classifying} derive four different types of trends, using 15 features to make their distinction. They distinguish trends triggered by news, current events, memes or commemorative tweets. Lehmann et al. \cite{lehmann2012dynamical} study different patterns of hashtag trends in \emph{Twitter}. They also observe four different classes of hashtag trends. 
Rattanaritnont et al. \cite{rattanaritnont2011study} propose to distinguish topics based on four factors, which are cascade ratio, tweet ratio, time of tweet and patterns in topic-sensitive hashtags.

We extend the model of \cite{bhamidi2012twitter} 
by mathematically describing the growth of a complete retweet graph. Our proposed model has two more parameters that define the shape of the resulting graph, in particular, the size and the density of its largest connected component. To the best of our knowledge, this is the first attempt to classify trends using a random graph model rather than algorithmic techniques or machine learning. The advantage of this approach is that it gives insight in emergence of the trend, which, in turn, is important for understanding and predicting the potential impact of social media on real world events.

\section{Datasets}\label{sec:data}
We use datasets containing tweets that have been acquired either using the \emph{Twitter} Streaming API\footnote{\url{https://dev.twitter.com/docs/streaming-apis}} or the \emph{Twitter} REST API\footnote{\url{https://dev.twitter.com/docs/api/1.1}}. Using the REST API one can obtain tweets or users from \emph{Twitter}'s databases. 
The Streaming API filters tweets that \emph{Twitter} parses during a day, for example, based on users, locations, hashtags, or keywords.

Most of the datasets used in this study were scraped by RTreporter, a company that uses an incoming stream of Dutch tweets to detect news for news agencies in the Netherlands. These tweets are scraped based on keywords, using the Streaming API. For this research, we selected several events that happened in the period of data collection, based on the wikipedia overviews of 2013 and 2014\footnote{\url{http://nl.wikipedia.org/wiki/2014} \& \url{http://nl.wikipedia.org/wiki/2013}}. We have also used two datasets scraped by TNO - Netherlands Organisation for Applied Scientific Research. The \emph{Project X} dataset contains tweets related to large riots in Haren, the Netherlands. This dataset is acquired by \emph{Twitcident}\footnote{\url{www.twitcident.com}}. For this study, we have filtered this dataset on two most important hashtags: \emph{\#projectx} and \emph{\#projectxharen}. The \emph{Turkish-Kurdish} dataset is described in more detail in Bouma et al.~\cite{bouma2012early}. A complete overview of the datasets, including the events and the keywords, is given in Table~\ref{tab:datasets}.  The size and the timespans for each dataset are given in Table~\ref{tab:data_stats}.

\begin{table}
\renewcommand\thetable{1}
\centering
{\scriptsize
\begin{longtable}{|l|l|l|}
\hline
&dataset&keywords\\
\hline
PX&Project X Haren& projectx, projectxharen\\\hline
TK&Demonstrations in Amsterdam &koerden, turken, rellen, museumplein,\\&related to the Turkish-Kurdish conflict& amsterdam\\\hline
WCS&World cup speedskating single distanced 2013& wkafstanden, sochi, sotsji\\\hline
W-A&Crowning of His Majesty King&troonswisseling, troon, Willem-Alexander, \\&Willem-Alexander in the Netherlands& Wim-Lex, Beatrix, koning, koningin\\\hline
ESF&Eurovision Song Festival&esf, Eurovisie Songfestival, ESF, \\&& songfestival, eurovisie\\\hline
CL&Champions Leage final 2013&Bayern Munchen, Borussia Dortmund, \\&&dorbay, borussia, bayern, borbay, CL\\\hline
Morsi&Morsi deposited as Egyption president&Morsi, afgezet, Egypte\\\hline
Train&Train crash in Santiago, Spain&Treincrash, treincrash, Santiago, \\&& Spanje, Santiago de Compostella, trein\\\hline
Heat&Heat wave in the Netherlands&hittegolf, Nederland\\\hline
Damascus&Sarin attack in Damascus&Sarin, Damascus, Syri\"{e}, syri\"{e}\\\hline
Peshawar&Bombing in Peshawar&Peshawar, kerk, zelfmoordaanslag, Pakistan\\\hline
Hawk&Hawk spotted in the Netherlands&sperweruil, Zwolle\\\hline
Pile-up&Multiple pile-ups in Belgium on the A19&A19, Ieper, Kortrijk, kettingbotsing\\\hline
Schumi&Michael Schumachar has a skiing accident&Michael Schumacher, ski-ongeval\\\hline
UKR&Rebellion in Ukrain&Azarov, Euromaidan, Euromajdan, Oekra\"{i}ne,\\&& opstand\\\hline
NAM&Treaty between NAM and Dutch government&Loppersum, gasakkoord, NAM, Groningen\\\hline
WCD&Michael van Gerwen wins PDC WC Darts&van Gerwen, PDC, WK Darts\\\hline
NSS&Nuclear Security Summit 2014&NSS2014, NSS,\\&& Nuclear Security Summit 2014,\\&& Den Haag\\\hline
MH730&Flight MH730 disappears&MH730, Malaysia Airlines\\\hline
Crimea&Crimea referendum for independance&Krim, referendum, onafhankelijkheid\\\hline
Kingsday&First Kingsday in the Netherlands&koningsdag, kingsday, koningsdag\\\hline
Volkert&Volkert van der Graaf released from prison&Volkert, volkertvandergraaf,\\&& Volkert van der Graaf\\\hline
\end{longtable}}
\medskip
\caption{Datasets: events and keywords (some keywords are in Dutch).
\label{tab:datasets}
}
\end{table}

\begin{table}[!ht]
\renewcommand\thetable{2}
\centering
{\scriptsize
\begin{longtable}{|l|c|c|c|r|r|}
\hline
dataset & year& first tweet & last tweet & \# tweets & \# retweets \\
\hline
PX & 2012& Sep 17 09:37:18 & Sep 26 02:31:15 & 31,144 & 15,357 \\
TK & 2011& Oct 19 14:03:23 & Oct 27 08:42:18 & 6,099 & 999 \\
WCS &2013&  Mar 21 09:19:06 & Mar 25 08:45:50 & 2,182 & 311 \\
W-A &2013& Apr 27 22:59:59 & May 02 22:59:25 & 352,157 & 88,594 \\
ESF & 2013& May 13 23:00:08 & May 18 22:59:59 & 318,652 & 82,968 \\
CL & 2013& May 22 23:00:04 & May 26 22:59:54 & 163,612 & 54,471 \\
Morsi &2013& Jun 30 23:00:00 & Jul 04 22:59:23 & 40,737 & 13,098 \\
Train & 2013& Jul 23 23:00:02 & Jul 30 22:59:41 & 113,375 & 26,534 \\
Heat & 2013& Jul 10 19:44:35 & Jul 29 22:59:58 & 173,286 & 42,835 \\
Damascus & 2013& Aug 20 23:01:57 & Aug 31 22:59:54 & 39,377 & 11,492 \\
Peshawar &2013& Sep 21 23:00:00 & Sep 24 22:59:59 & 18,242 & 5,323 \\
Hawk & 2013& Nov 11 23:00:07 & Nov 30 22:58:59 & 54,970 & 19,817 \\
Pile-up &2013&  Dec 02 23:00:15 & Dec 04 22:59:57 & 6,157 & 2,254 \\
Schumi &2013-14& Dec 29 02:43:16 & Jan 01 22:54:50 & 13,011 & 5,661 \\
UKR &2014& Jan 26 23:00:36 & Jan 31 22:57:12 & 4,249 & 1,724 \\
NAM &2014& Jan 16 23:00:22 & Jan 20 22:59:49  & 41,486 & 14,699 \\
WCD &2013-14& Dec 31 23:03:48 & Jan 02 22:59:05 & 15,268 & 5,900 \\
NSS &2014& Mar 23 23:00:06 & Mar 24 22:59:56 & 29,175 & 13,042 \\
MH730 &2014& Mar 08 00:18:32 & Mar 28 22:40:44 & 36,765 & 17,940 \\
Crimea &2014& Mar 13 23:02:22 & Mar 17 22:59:57 & 18,750 & 5,881 \\
Kingsday &2014& Apr 26 23:00:00 & Apr 29 22:53:00 & 7,576 & 2,144 \\
Volkert &2014& Apr 30 23:08:14 & May 04 22:57:06 & 9,659 & 4,214 \\\hline
\end{longtable}}
\medskip
\caption{Characteristics of the datasets.\label{tab:data_stats}}
\end{table}

\begin{figure}[!ht]
\centering
\includegraphics[height=0.6\textwidth,angle=270]{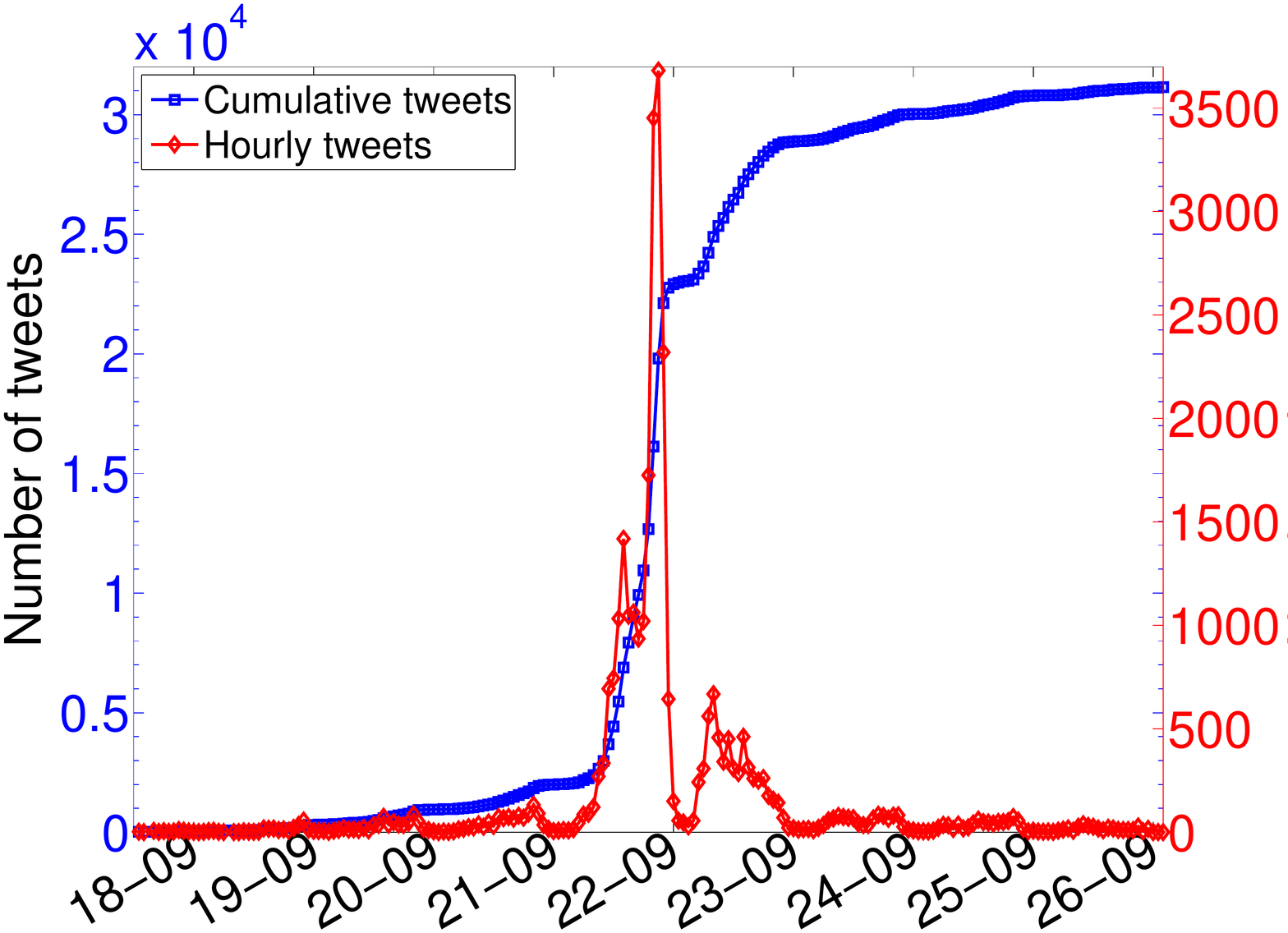}
\caption{{\it Project X} Number of tweets and cumulative number of tweets per hour.} \label{fig:prog_tweet_sel}

\end{figure}

For each dataset we have observed there is at least one large peak in the progression of the number of tweets. For example, Figure~\ref{fig:prog_tweet_sel} shows such peak in {\it Twitter} activity for the {\it Project X} dataset.

When a retweet is placed on \emph{Twitter}, the Streaming API returns the retweet together with the message that has been retweeted. We use this information to construct the retweet trees of every message and the user IDs for each posted message. The tweet and graph analysis is done using \emph{Python} and its modules \emph{Tweepy}\footnote{\url{http://www.tweepy.org/}} and \emph{NetworkX}\footnote{\url{http://networkx.github.io/}}. In this paper, we investigate the dynamics of retweet graphs with the goal to predict peaks in {\it Twitter} activity and classify the nature of trends.

\section{Retweet graphs}
\label{sec:retweet graphs}

Our main object of study is the retweet graph $G=(V,E)$, which is a graph of users that have participated in the discussion on a specific topic. A directed edge $e=(u,v)$ indicates that user $v$ has retweeted  a tweet of $u$. We observe the retweet graph at the time instances $t=0,1,2,\ldots$, where either a new node or a new edge was added to the graph, and we denote by $G_t=(V_t,E_t)$ the retweet graph at time $t$. As usual, the out- (in-) degree of node $u$ is the number of directed edges with source (destination) in $u$. In what follows, we model and analyse the properties of  $G_t$. For every new message initiated by a new user $u$ a tree $T_{u}$ is formed. Then, ${\cal T}_t$ denotes the forest of message trees. Note that in our model a new message from an already existing user $u$ (that is, $u\in {\cal T}_t$) does not initiate a new message tree. We define $|{\cal T}_t|$ as the number of new users that have started a message tree up to time $t$. 

\begin{figure}[!ht]
\vspace{-25pt}
\centering
\subfloat[Edge density.\label{fig:prog_edges_nodes}]{
\includegraphics[height=0.5\textwidth,angle=270]{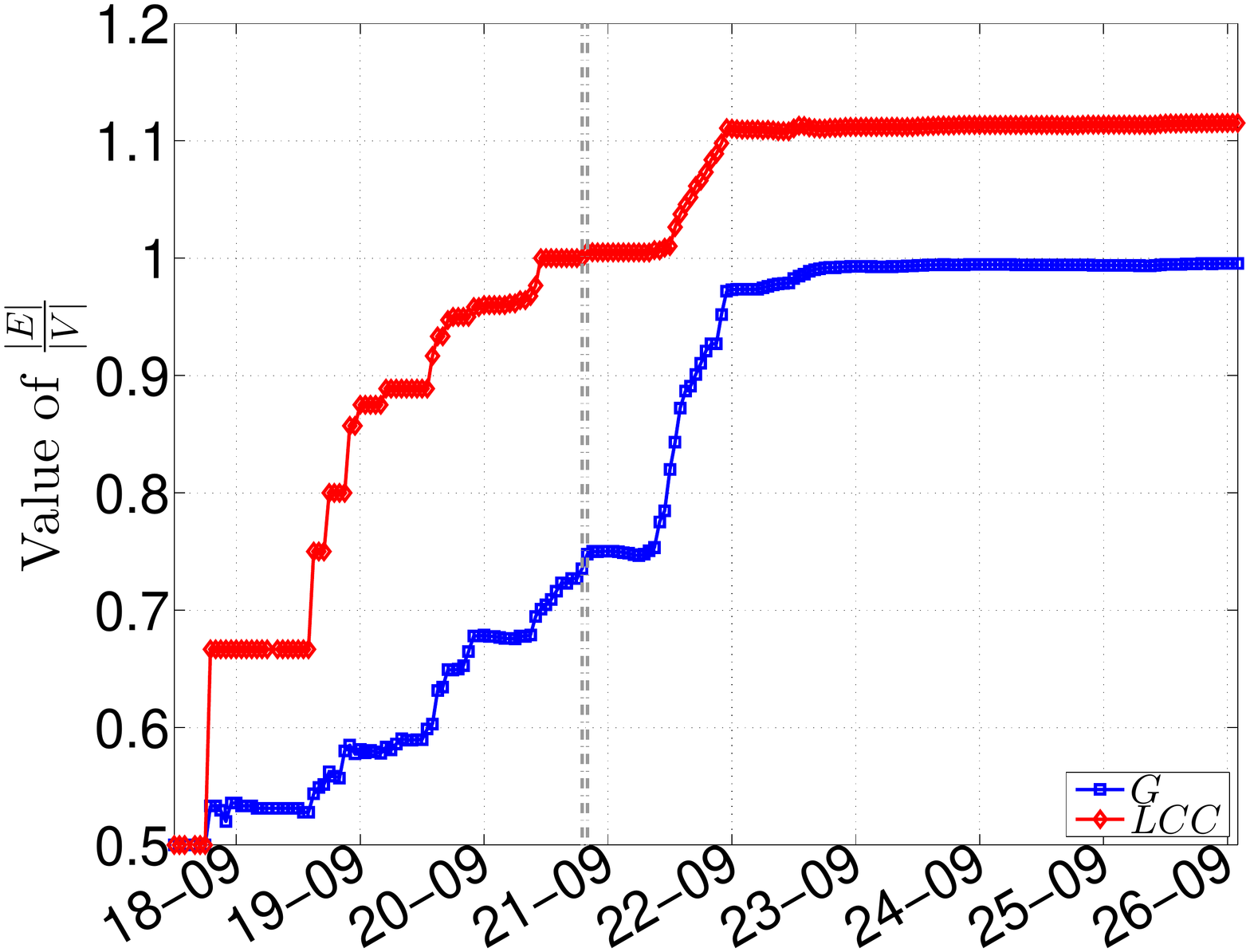}}
\subfloat[Size of LCC.\label{fig:prog_gc}]{
\includegraphics[height=0.5\textwidth,angle=270]{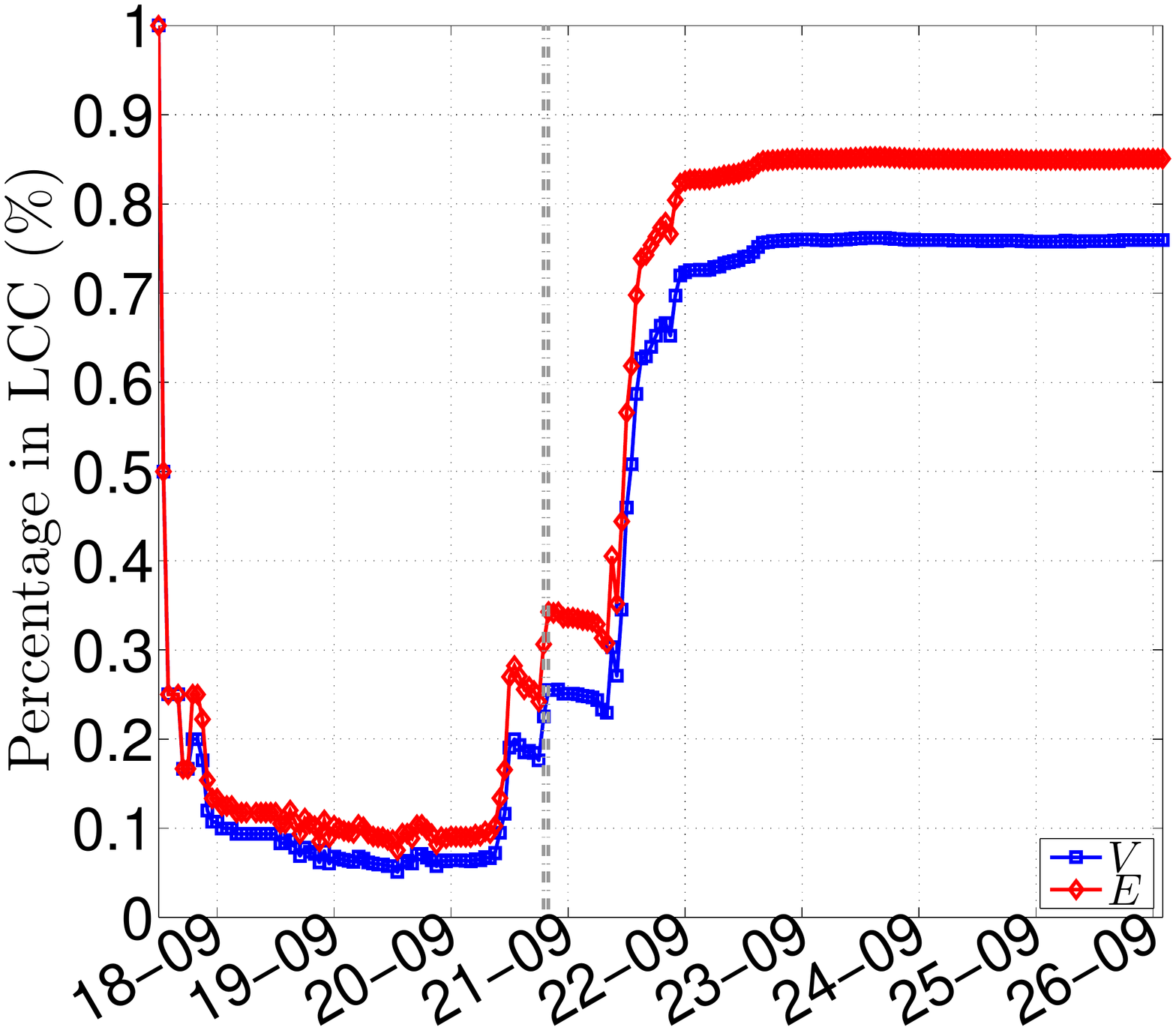}}
\caption{Progression for the edge density (a) and the size of the LCC (b) in the \emph{Project X} dataset.\label{fig:PX}}

\end{figure}

After analyzing multiple characteristics of the retweet graphs for every hour of their progression, we found that the size of the largest (weakly) connected component (LCC) and its density are the most informative characteristics for predicting the peak in {\it Twitter}. In Figure~\ref{fig:PX} we show the development of these characteristics in the \emph{Project X} dataset. One day before the actual event, we observe a very interesting phenomenon in the development of the edge density of the LCC in Figure~\ref{fig:prog_edges_nodes}. Namely, at some point the edge density of the LCC exceeds $1$ (indicated by the dash-dotted gray lines), i.e. there is more than one retweet per user on average. We shall refer to this as the \emph{densification} (or dens.) of the LCC. Furthermore, the relative size of the LCC increases from 18\% to 25\% as well, see Figure \ref{fig:prog_gc}.

We have observed a densification of the LCC in each dataset that we have studied. Indeed, when the LCC grows its density must become at least one (each node is added to the LCC together with at least one edge). However, we have also observed that in each dataset the densification occurs before the main peak, but the scale of densification is different. For example, in the {\it Project X} dataset the densification already occurs one day before the peak activity. Plausibly, in this discussion, that ended up in riots, a group of people was actively participating before the event. On the other hand, in the {\it WCS} dataset, which tweets about an ongoing sport event,  the densication of the LCC occurs during the largest peak. This is the third peak in the progression. Hence, our experiments suggest that the time of densification has predictive value for trend progression and classification. See Table~\ref{tab:param_values} for the density of the LCC in each dataset at the end of the progression.

\section{Model}\label{sec:model}
Our goal is to design a model that captures the development of trending behaviour. In particular, we need to capture the phenomenon that disjoint components of the retweet graph join together forming the largest component, of which the density of edges may become larger than one. To this end, we employ the superstar model of Bhamidi et al. \cite{bhamidi2012twitter} for modelling distinct components of the retweet graph, and add the mechanism for new components to arrive and the existing  components to merge. For the sake of simplicity of the model we neglect the friend-follower network of \emph{Twitter}. Note that in \emph{Twitter} every user can retweet any message sent by any public user, which supports our simplification. 

At the start of the progression, we have the graph $G_0$. In the analysis of this section, we assume that $G_0$ consists of a single node. Note that in reality, this does not need to be the case: any directed graph can be used as an input graph $G_0$. In fact, in Section~\ref{sec:practice} we start with the actual retweet graph at a given point in time, and then use the model to build the graph further to its final size. 

We consider the evolution of the retweet graph in time $(G_t)_{t\ge 0}$.  We use a subscript $t$ to indicate $G_t$ and related notions at time $t$. We omit the index $t$ when referring to the graph at the end of the progression. 

Recall that $G_t$ is a graph of {\it users}, and an edge $(u,v)$ means that $v$ has retweeted a tweet of $u$. We consider  time instances $t=1,2,\ldots$ when either a new node or a new edge is added to the graph $G_{t-1}$. We distinguish three types of changes in the retweet graph:

\begin{itemize}
\item[$\circ$] $T1$: a new user $u$ has posted a new message on the topic, node $u$ is added to $G_{t-1}$;
\item[$\circ$] $T2$: a new user $v$ has retweeted an existing user $u$,  node $v$ and edge $(u,v)$ are added to $G_{t-1}$;
\item[$\circ$] $T3$: an existing user $v$ has retweeted another existing user $u$, edge $(u,v)$ is added to $G_{t-1}$.
\end{itemize}

The initial node is equivalent to a $T1$ arrival at time $t=0$. Assume that each change in $G_t$ at $t=1,2,\ldots$ is $T1$ with probability $\lambda/(1+\lambda)$, independently of the past. Also, assume that a new edge (retweet) is coming from a new user with probability $p$. Then the probabilities of $T1$, $T2$ and $T3$ arrivals are, respectively $\frac{\lambda}{\lambda+1}$, $\frac{p}{\lambda+1}$, $\frac{1-p}{\lambda+1}$. The parameter $p$ is governing the process of components merging together, while $\lambda$ is governing the  arrival of new components in the graph. 

For both $T2$ and $T3$ arrivals we define the same mechanism for choosing the source of the new edge $(u,v)$ as follows. 

Let $u_0,u_1,\ldots$ be the users that have been added to the graph as $T1$ arrivals, where $u_0$ is the initial node. Denote by $T_{i,t}$ the subgraph of $G_t$ that includes $u_i$ and all users that have retweeted the message of $u_i$ in the interval $(0,t]$. We call such a subgraph a message tree with root $u_i$.  We assume that the probability that a $T2$ or $T3$ arrival at time $t$ will attach an edge to one of the nodes in $T_{i,t-1}$ with probability $p_{T_{i,t-1}}$, proportional to the size of the message tree:
\[ p_{T_{i,t-1}} = \frac{|T_{i,t-1}|}{\sum_{T_{j,t-1}\subset{\cal T}_{t-1}}|T_{j,t-1}|}. \]
This creates a preferential attachment mechanism in the formation of the message trees. Next, a node in the selected message  tree $T_{i,t-1}$ is chosen as the source node following the superstar attachment scheme \cite{bhamidi2012twitter}: with probability $q$, the new retweet is attached to $u_i$, and with probability $1-q$, the new retweet is attached to any other vertex, proportional to the preferential attachment function of the node, that we choose to be the number of children of the node plus one. 

Thus we employ the superstar-model, which was suggested in \cite{bhamidi2012twitter} for modelling the largest connected component of the retweet graph on a given topic, in order to describe a progression mechanism for a single retweet tree. Our extensions compared to \cite{bhamidi2012twitter} are that we allow new message trees to appear ($T1$ arrivals), and that different message trees may either remain disconnected or get connected by a $T3$ arrival.  

For a $T3$ arrival, the target of the new edge $(u,v)$ is chosen uniformly at random from $V_{t-1}$, with the exception of the earlier chosen source node $u$, to prevent self-loops. That is, any user is equally likely to retweet a message from another existing user.

Note that, in our setting, it is easy to introduce a different superstar parameter $q_{T_i}$ for every message tree $T_i$. This way one could easily implement specific properties of the user that starts the message tree, e.g. his/her number of followers. For the sake of simplicity, we choose the same value of $q$ for all message trees. Also note that we do not include tweets and retweets that do not result in new nodes or edges in a retweet graph. This could be done, for example, by introducing dynamic weights of vertices and edges, that increase with new tweets and retweets. Here we consider only an unweighted model.

\subsection{Growth of the graph}
The average degree, or edge density, is one of the aspects through which we give insight to the growth of the graph. The essential properties of this characteristic are presented in Theorem~\ref{thm:exp_av_degree}. The proof is given in the Appendix. 
\begin{thm} Let $\tau_n$ be the time when node $n$ is added to the graph. Then
\begin{eqnarray}
\E{\frac{|E_{\tau_n}|}{|V_{\tau_n}|}} &=& \frac{1}{\lambda+p}-\frac{1}{n(\lambda+p)},\label{eq:exp_av_degree}
\\
\var{\frac{|E_{\tau_n}|}{|V_{\tau_n}|}} &=& \frac{(n-1)(\lambda +1- p)}{n^2(\lambda+p)^2}.
\label{eq:var_av_degree}
\end{eqnarray}
\label{thm:exp_av_degree}
\end{thm}

Note that the variance of the average degree in (\ref{eq:var_av_degree}) converges to zero as $n\to \infty$ at rate $\frac{1}{n}$. 

The next theorem studies the observed ratio between $T2$ and $T3$ arrivals (new edges) and $T1$ arrivals (new nodes with a new message). As we see from the theorem, this ratio can be used for estimating the parameter $\lambda$. The proof is given in the Appendix.

\begin{thm} Let $G_t=(V_t,E_t)$ be the retweet graph at time $t$, let ${\cal T}_t$ be the set of all message trees in $G_t$.  Then
\begin{eqnarray}
&&\E{\frac{|E_t|}{|{\cal T}_t|}} = \lambda^{-1}\cdot\left(1-\left(\frac{1}{\lambda+1}\right)^t\right),\label{eq:exp_edges_discussion}
\\
&&\lim_{t\to\infty}\frac{\lambda^3 t}{(1+\lambda)^2}\var{\frac{|E_t|}{|{\cal T}_t|}} =1, 
\label{eq:var_edges_discussion}
\end{eqnarray}
Furthermore, 
\begin{equation}
\label{eq:distr_edges_discussion}
\frac{\lambda^{3/2}\sqrt{t}}{\lambda+1}\left(\frac{|E_t|}{|{\cal T}_t|}-\frac{1}{\lambda}\right)\stackrel{D}{\to} Z,
\end{equation}
where $Z$ is a standard normal $ N(0,1)$ random variable, and  $\stackrel{D}{\to}$ denotes convergence in distribution.
\label{thm:exp_edges_discussion}
\end{thm}

Note that, as expected from the definition of $\lambda$,
\begin{equation}
\lim_{t\rightarrow\infty}\E{\frac{|E_t|}{|{\cal T}_t|}} = \lambda^{-1}.\label{eq:lim_exp_edges_discussion}
\end{equation} 
This will be used in Section \ref{sec:practice} for estimating $\lambda$.

\subsection{Component size distribution}
{\sloppy In the following, we assume that $G_t$ consists of $m$ connected components ($C_1,C_2,\ldots,C_m$) with known respective sizes ($|C_1|,\ldots,|C_m|$). We aim to derive expressions for the distribution of the component sizes in $G_{t+1}$. 

}
\begin{lem} The distribution of the sizes of the components of $G_{t+1}$, given $G_t$ is as follows,
\begin{equation}
\begin{cases}
|C_1|,\ldots,|C_i|,|C_j|,\ldots,|C_m|,1 & \mbox{w.p. }\frac{\lambda}{\lambda+1}\\
|C_1|,\ldots,|C_i|+1,|C_j|,\ldots,|C_m| & \mbox{w.p. }\frac{p}{\lambda+1}\cdot\frac{|C_i|}{|V|}\\
|C_1|,\ldots,|C_i|+|C_j|,\ldots,|C_m| & \mbox{w.p. }\frac{1-p}{\lambda+1}\cdot\frac{2\cdot|C_i|\cdot|C_j|}{|V|^2-|V|}\\
|C_1|,\ldots,|C_i|,|C_j|,\ldots,|C_m| & \mbox{w.p. }\frac{1-p}{\lambda+1}\cdot\frac{\sum_{k=1}^m|C_k|^2-|C_k|}{|V|^2-|V|}
\end{cases}
\end{equation}
\label{lem:distribution_component_sizes_one_step}\end{lem}

The proof of Lemma~\ref{lem:distribution_component_sizes_one_step} is given in the Appendix. Lemma \ref{lem:distribution_component_sizes_one_step} provides a recursion for computing the distribution of component sizes. However, the computations are highly demanding if not infeasible. Also, deriving an exact expression of the distribution of the component sizes at time $t$ is very cumbersome because they are hard and they strongly depend on the events that occurred at $t=0,\ldots,t-1$. Note that if $p=1$, there is a direct correspondence between our model and the infinite generalized P\'{o}lya process~\cite{chung2003generalizations}. However, this case is uninformative as there are no $T3$ arrivals. Therefore, in the next section we resort to simulations to investigate the sensitivity of the graph characteristics to the model parameters.

\subsection{Influence of $q$, $p$ and $\lambda$}
We analyze the influence of the model parameters $\lambda$, $p$ and $q$ on the characteristics of the resulting graph numerically using simulations. To this end, we fix two out of three parameters and execute multiple simulation runs of the model, varying the values for the third parameter. We start simulations with graph $G_0$, consisting of one node. We perform $50$ simulation runs for every parameter setting and obtain the average values over the individual runs for given parameters.  

Parameter $q$ affects the degree distribution~\cite{bhamidi2012twitter} and the overall structure of the graph. If $q=0$, then the graph contains less nodes that have many retweets. If $q=1$ each edge is connected to a superstar, and the graph consists of star-like sub graphs, some of which are connected to each other. In the {\it Project X} dataset, which is our main case study, $q\approx0.9$ results in a degree distribution that closely approximates the data. Since degree distributions are not in the scope of this paper, we omit these results for brevity. 

\begin{figure}[!ht]
\vspace{-25pt}
\subfloat[$\frac{|E_{\mbox{\tiny{LCC}}}|}{|V_{\mbox{\tiny{LCC}}}|}$\label{fig:E_over_V_in_GC_numerical}]{
\includegraphics[width=0.52\textwidth]{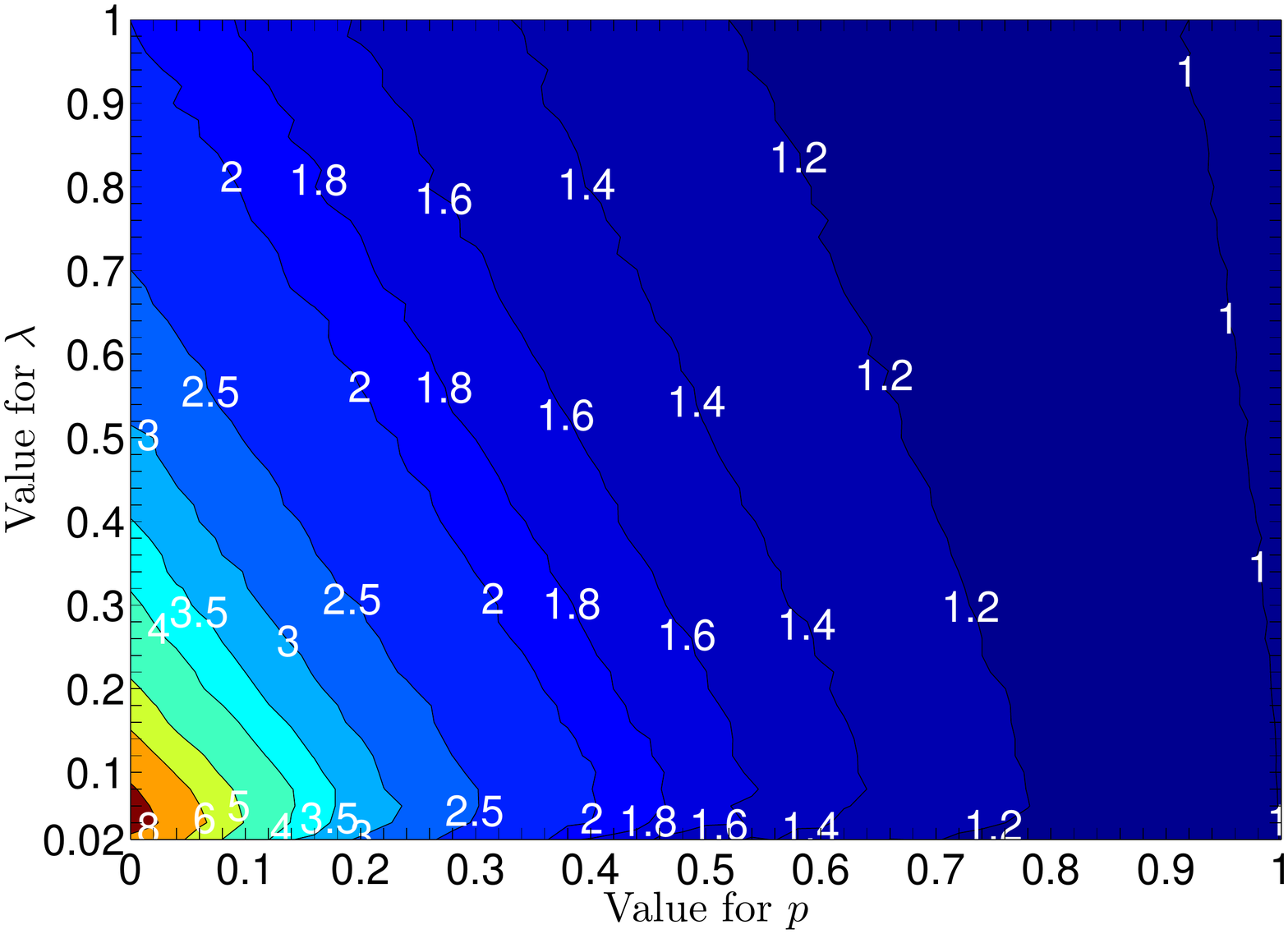}}
\subfloat[$\frac{|V_{\mbox{\tiny{LCC}}}|}{|V|}$\label{fig:frac_in_GC_numerical}]{
\includegraphics[width=0.52\textwidth]{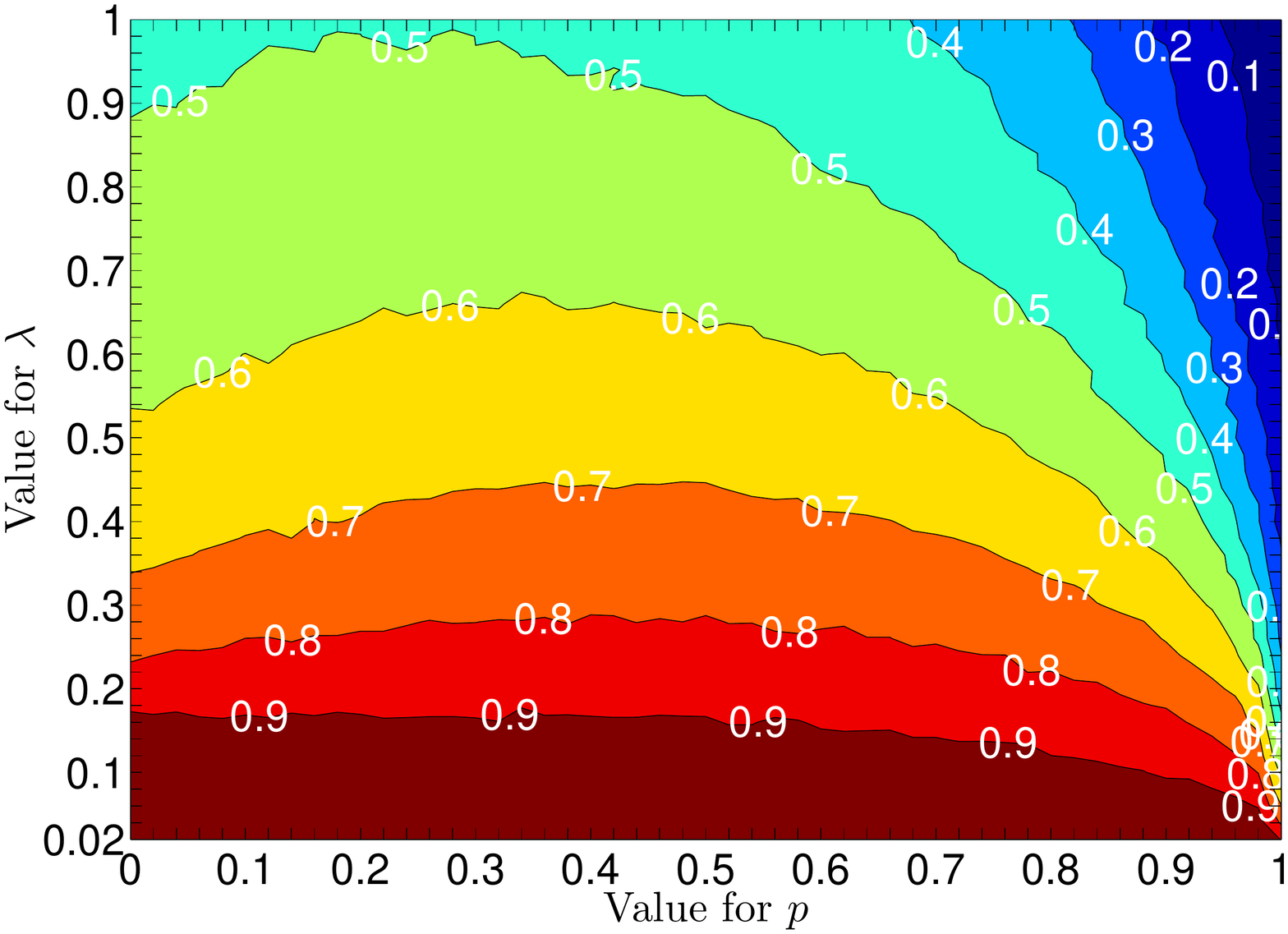}}
\caption{Numerical results for the model using $q=0.9$ and $t=1,{}000$\label{fig:numerical}.}
\vspace{-20pt}
\end{figure}

We compare the results for two measures that produced especially important characteristics of the \emph{Project X} dataset:  $\frac{|E_{\mbox{\tiny{LCC}}}|}{|V_{\mbox{\tiny{LCC}}}|}$ and $\frac{|V_{\mbox{\tiny{LCC}}}|}{|V|}$. These characteristics do not depend on $q$. In simulations, we set $t=1,{}000$, $q=0.9$ and vary the values for $p$ and $\lambda$. the results are give in Figure~\ref{fig:numerical}. 

We see that the edge density in the LCC  in Figure \ref{fig:E_over_V_in_GC_numerical} decreases with $\lambda$ and $p$. Note that according to (\ref{eq:exp_av_degree}),  $|E|/|V|$ is well approximated by $1/(\lambda+p)$ when $\lambda$ or $p$ are large enough. The edge density in LCC shows a similar pattern,  but it is slightly higher than in the whole graph. When $\lambda$ and $p$ are small, there are many $T3$ arrivals, and new nodes are not added frequently enough. This results in an unexpected non-monotonic behaviour of the edge density near the origin. For the fraction of nodes in the LCC, depicted in Figure \ref{fig:frac_in_GC_numerical}, we see that the parameter $\lambda$ is most influential. The parameter $p$ is of considerable influence only when it is large.

\section{The model in practice}\label{sec:practice}
In this section we obtain parameter estimators for our model and compare the model to the datasets discussed in Section \ref{sec:data}.

Using Theorem \ref{thm:exp_edges_discussion}, we know that $\frac{|E_t|}{|{\cal T}_t|}$ converges to $\lambda^{-1}$ as $t\rightarrow\infty$. Thus, we suggest the following estimator for $\lambda$ at time $t> 0$:
\begin{equation}
\hat{\lambda}_t = \frac{|{\cal T}_t|}{|E_t|}.\label{eq:estimate_lambda}
\end{equation}

Second, we derive an expression for $\hat{p}_t$ using (\ref{eq:exp_av_degree}) and substituting (\ref{eq:estimate_lambda}) for $\lambda$:
\begin{eqnarray}
\hat{p}_t & = & \frac{|V_t|-|{\cal T}_t|-1}{|E_t|}.
\label{eq:estimate_p}
\end{eqnarray}

Since the \emph{Twitter} API only gives back the original message of a retweet and not the level in the progression tree of that retweet, we can not determine $q$ easily from the data. Since this parameter does not have a large influence on the outcomes of the simulations, we choose this parameter to be $0.9$ for all datasets.

Notice that we can obtain the numbers ($|E_t|$, $|{\cal T}_t|$ and $|V_t|$) directly from a given retweet graph for each $t=1,2,\ldots$. The computed estimators for our datasets are displayed in Table~\ref{tab:param_values}. 

\begin{table}[!ht]
\vspace{-10pt}
\centering{\scriptsize
\begin{tabular}{c|cc|ccc|ccc}
 &&& \multicolumn{3}{c|}{actual progression} & \multicolumn{3}{c}{simulations (starting at dens.)} \\
dataset& $\hat{\lambda}$ & $\hat{p}$ & $\frac{|V_{LCC}|}{|V|}$ & $\frac{|E|}{|V|}$ & $\frac{|E_{LCC}|}{|V_{LCC}|}$ & $\frac{|V_{LCC}|}{|V|}$ & $\frac{|E|}{|V|}$ & $\frac{|E_{LCC}|}{|V_{LCC}|}$ \\
\hline
PX & .23 & .78 & .76 & 1.00 & 1.12 & .54 & .75 & 1.08\\
TK & .42 & .85 & .25 & .79 & 1.00 & .54 & .74 & 1.08 \\
WCS & .49 & .73 & .20 & .81 & .99 & .49 & .95 & 1.90 \\
W-A & .41 & .52 & .67 & 1.07 & 1.30 & .40 & .62 & 1.41 \\
ESF & .38 & .43 & .73 & 1.24 & 1.48 & .45 & .69 & 1.42 \\
CL & .40 & .72 & .44 & .90 & 1.22 & .46 & .66 & 1.16 \\
Morsi & .60 & .55 & .39 & .87 & 1.20 & .47 & .67 & 1.17 \\
Train & .54 & .78 & .28 & .76 & 1.04 & .50 & .70 & 1.17 \\
Heat & .42 & .59 & .60 & .99 & 1.23 & .41 & .72 & 1.68 \\
Damascus & .58 & .51 & .46 & .92 & 1.24 & .44 & .65 & 1.30 \\
Peshawar & .54 & .68 & .31 & .82 & 1.18 & .53 & .75 & 1.25 \\
Hawk & .38 & .38 & .82 & 1.31 & 1.45 & .49 & .76 & 1.43 \\
Pile-up & .33 & .64 & .65 & 1.03 & 1.24 & .58 & .93 & 1.54 \\
Schumi & .38 & .83 & .33 & .82 & 1.08 & .56 & .77 & 1.07 \\
UKR & .72 & .37 & .53 & .91 & 1.12 & .50 & .75 & 1.38 \\
NAM & .44 & .48 & .50 & 1.09 & 1.51 & .45 & .72 & 1.51 \\
WCD & .26 & .81 & .66 & .94 & 1.10 & .64 & .83 & 1.07 \\
NSS & .26 & .62 & .79 & 1.13 & 1.26 & .23 & .35 & 1.21 \\
MH730 & .33 & .52 & .15 & 1.18 & 1.00 & .56 & .76 & 1.09 \\
Crimea & .44 & .63 & .51 & .93 & 1.19 & .52 & .72 & 1.12\\
Kingsday & .47 & .92 & .07 & .72 & 1.11 & .47 & .67 & 1.15\\
Volkert & .29 & .55 & .79 & 1.18 & 1.31 & .64 & .87 & 1.22\\ 
\end{tabular}}
\medskip
\caption{Estimated parameter values using complete dataset, simulation and progression properties.\label{tab:param_values}}
\end{table}

Next, we compare 50 simulations of the datasets from the point of densification of the LCC until the graph has reached the same size as the actual dataset. We display the average outcomes of these simulations and compare them to the actual properties of the retweet graphs of each dataset in Table~\ref{tab:param_values}.

Here we see diverse results per dataset in the simulations. For the \emph{CL, Morsi} and \emph{WCD} datasets, the simulations are very similar to the actual progressions. However, for some datasets, for instance the \emph{ESF} dataset, simulations are far off. In general, the model predicts the density of the LCC quite well for many datasets, but tends to overestimate the size of the LCC. We notice that current random graph models for networks usually capture one or two essential features, such as degree distribution, self-similarity, clustering coefficient or diameter. Our model captures both degree distribution and, in many cases, the density of the LCC. It seems that our model performs better on the datasets that have a singular peak rather than a series of peaks. We have observed on the data that each peak activity has a large impact on the parameters estimation. We will strive to adopt the model for incorporating different rules for activity during peaks, and improving results on the size of the LCC. 

\section{Conclusion and Discussion}\label{sec:conclusion}
We have found that our model performs well in modelling the retweet graph for tweets regarding a singular topic. However, there is a room for improvement when the dataset covers a prolonged discussion with  users activity fluctuating over time. 

A possible extension of the present work is incorporating more explicitly the time aspect into our model. We could for example add the notion of `novelty', like G\'{o}mez et al. in \cite{gomez2012likelihood}, taking into account that e.g. the retweet probability for a user may decrease the longer he/she remains silent after having received a tweet. But also other model parameters may be assumed to vary over time. In addition, we propose to analyse  the clustering coefficient of a node in the network model and, in particular, to investigate how it evolves over time. This measure (see \cite{watts1998collective}) provides more detailed insight in how the graph becomes denser, making it possible to distinguish between local and global density.

%

\bibliographystyle{abbrv}
\bibliography{biblio}

\section*{Appendix}
\appendix
\subsection*{A1. Proof of Theorem~\ref{thm:exp_av_degree}}
\begin{proof} The proof is based on the fact that  the total number of edges $|E_{\tau_n}|$ equals a total number of the $T2$ and $T3$ arrivals on $(0,\tau_n]$. By definition, $(0,\tau_n]$ contains exactly $(n-1)$ of $T1$ or $T2$ arrivals, hence, the number of $T2$ arrivals has a Binomial distribution with number of trials equal to $(n-1)$, and success probability $P(T2\,|\mbox{ $T1$ or  $T2$})=\frac{p}{\lambda+p}$.
Next, the number of $T3$ arrivals on $[\tau_i,\tau_{i+1})$, where $i=1,\ldots,n-1$, has a shifted geometric distribution, namely, the probability of $k$ $T3$ arrivals on $[\tau_i,\tau_{i+1})$ is \[\left(1-\frac{1-p}{\lambda+1}\right)\left(\frac{1-p}{\lambda+1}\right)^k,\; k=0,1,\ldots.\] Observe that there have been $n-1$ of these transitions from 1 node to $n$. Hence,
the number of $T3$ arrivals on $(0,\tau_n]$ is the sum of $(n-1)$ i.i.d. Geometric random variables with mean $\frac{1-p}{\lambda+p}$. Summarizing the above, we obtain (\ref{eq:exp_av_degree}). For (\ref{eq:var_av_degree}) we also need to observe that the number of $T2$ and $T3$ arrivals on $[0,\tau_n]$ are independent.
\end{proof}

\subsection*{A2. Proof of Theorem~\ref{thm:exp_edges_discussion}}
\begin{proof}
Let $X_t$ be the number of $T2$ and $T3$ arrivals by time $t$. Note that $|E_t|=X_t$, and $|{\cal T}_t|=t-X_t+1$, which is the number of $T_1$ arrivals on $[0,t]$, since the first node at time $t=0$ is by definition a $T1$ arrival. Note that $X_t$  has a binomial distribution with parameters $t$ and $\prob{\mbox{$T2$ arrival}}+\prob{\mbox{$T3$ arrival}}=\frac{1}{\lambda+1}$. Furthermore, the number of $T1$ arrivals is $t-X_t+1$ since the first node at time $t=0$ is by definition a $T1$ arrival. Hence,
\begin{eqnarray*} 
\E{\frac{|E_t|}{|{\cal T}_t|}} & = & \sum_{i=1}^t\frac{i}{t-i+1}{t\choose i}\left(\frac{1}{\lambda+1}\right)^i\left(\frac{\lambda}{\lambda+1}\right)^{t-i}\\
& = & \frac{1}{\lambda}\cdot\sum_{i=1}^t{t \choose i-1}\left(\frac{1}{\lambda+1}\right)^{i-1}\left(\frac{\lambda}{\lambda+1}\right)^{t-i+1},\\
\end{eqnarray*}
which proves (\ref{eq:exp_edges_discussion}). Next, we write
\begin{eqnarray} 
&&\E{\left(\frac{|E_t|}{|{\cal T}_t|}\right)^2}  
=\sum_{i=0}^t\left(\frac{i}{t-i+1}\right)^2{t\choose i}\left(\frac{1}{\lambda+1}\right)^i\left(\frac{\lambda}{\lambda+1}\right)^{t-i}\nonumber\\
&&\quad =\frac{1}{\lambda}\cdot\sum_{i=1}^t\frac{i}{t-i+1}{t\choose i-1}\left(\frac{1}{\lambda+1}\right)^{i-1}\left(\frac{\lambda}{\lambda+1}\right)^{t-i+1}\nonumber\\
&&=\frac{1}{\lambda}\E{\frac{t+1}{t-X_t}\mathbbm{1}_{\{X_t\le t-1\}}}-\frac{1}{\lambda}\left(1-\left(\frac{1}{1+\lambda}\right)^t\right),
\label{eq:exp_edges_discussion2}
\end{eqnarray}
where $\mathbbm{1}_{\{A\}}$ is an indicator of event $A$. Denoting
\begin{equation}
\label{eq:z0}Z_t=\frac{X_t-\E{X_t}}{\sqrt{\var{X_t}}}=\frac{(\lambda+1)X_t-t}{\sqrt{\lambda t}},\end{equation}
we further write
\begin{eqnarray}
\label{eq:z1}
&&\E{\frac{t+1}{t-{X_t}}\mathbbm{1}_{\{X_t\le t-1\}}}=\E{\frac{(t+1)(\lambda+1)}{\lambda t(1-\frac{Z_t}{\sqrt{\lambda t}})}\mathbbm{1}_{\{Z_t\le \sqrt{\lambda }t-\frac{\lambda+1}{\sqrt{\lambda t}}\}}}.
\end{eqnarray}
 We now split the indicator above as follows:
 \begin{equation}
 \label{eq:z2}
\mathbbm{1}_{\{Z_t\le -\sqrt{\lambda }t\}}+  \mathbbm{1}_{\{-\sqrt{\lambda }t< Z_t<\sqrt{\lambda }t/2\}}+\mathbbm{1}_{\{\sqrt{\lambda }t/2\le Z_t\le \sqrt{\lambda }t-\frac{\lambda+1}{\sqrt{\lambda t}}\}}.
 \end{equation}
 For the first and the third term we use the Chernoff bound:
 \begin{eqnarray}
 \label{eq:z3} \E{\frac{1}{1-\frac{Z_t}{\sqrt{\lambda t}}}\mathbbm{1}_{\{Z_t\le -\sqrt{\lambda }t\}}}&\le& 2e^{-\lambda t/4},\\
 \E{\frac{1}{1-\frac{Z_t}{\sqrt{\lambda t}}}\mathbbm{1}_{\{\sqrt{\lambda }t/2\le Z_t\le \sqrt{\lambda }t-\frac{\lambda+1}{\sqrt{\lambda t}}\}}}&\le& \frac{\sqrt{\lambda t}}{\lambda+1}2e^{-\lambda t/16},
 \label{eq:z4}
  \end{eqnarray}
  and notice that both expressions above converge to zero faster than $1/t$. For the second case, note first that $\E{Z_t}=0$ and hence it follows from (\ref{eq:z0}) and (\ref{eq:z2})--(\ref{eq:z4}) that, as $t\to\infty$,
  \[\E{Z_t\mathbbm{1}_{\{-\sqrt{\lambda }t< Z_t<\sqrt{\lambda }t/2\}}}=o\left(\frac{1}{t}\right).\]
   Then we use the Taylor expansion to obtain:
\begin{eqnarray}
&&\left|\E{\frac{1}{1-\frac{Z_t}{\sqrt{\lambda t}}}\mathbbm{1}_{\{-\sqrt{\lambda }t< Z_t<\sqrt{\lambda }t/2\}}}-1\right|\nonumber\\
&&\hspace{2cm} \le \E{\frac{Z_t^2}{\lambda t}}+2\E{\frac{|Z_t|^3}{(\lambda t)^{3/2}}}+o\left(\frac{1}{t}\right),\label{eq:var_estimate}
\end{eqnarray}
as $t\to\infty$. By the central limit theorem, ${Z_t}\overset{D}{\longrightarrow}Z$ as $t\to\infty$.
Furthermore, for $r>0$, the convergence of moments holds~\cite{hall1983rate}: 
$\lim_{t\to\infty}\E{|{Z_t}|^r}=\E{|Z|^r}$. In particular, in (\ref{eq:var_estimate}),  $\E{|{Z_t}|^3}$ converges to a constant, and 
$\E{{Z_t}^2}$ converges to 1 as $t\to\infty$. Thus, using (\ref{eq:exp_edges_discussion2})--(\ref{eq:z1}) and (\ref{eq:exp_edges_discussion}) we write
\begin{eqnarray*}
&&\var{\frac{|E_t|}{|{\cal T}_t|}}=\E{\left(\frac{|E_t|}{|{\cal T}_t|}\right)^2} -\left(\E{\frac{|E_t|}{|{\cal T}_t|}}\right)^2\\
 &&=\E{\frac{(t+1)(\lambda+1)}{\lambda t(1-\frac{Z_t}{\sqrt{\lambda t}})}\mathbbm{1}_{\{Z_t\le \sqrt{\lambda }t-\frac{\lambda+1}{\sqrt{\lambda t}}\}}}-\frac{1}{\lambda}-\frac{1}{\lambda^2} +o\left(\frac{1}{t}\right).
\end{eqnarray*}
Now, subsequently using (\ref{eq:z2}) -- (\ref{eq:var_estimate}), we get
\begin{eqnarray*}
\var{\frac{|E_t|}{|{\cal T}_t|}}&=&
\frac{1}{\lambda}\,\frac{(t+1)(\lambda+1)}{\lambda t}\left(1+\frac{1}{\lambda t}+o\left(\frac{1}{t}\right)\right)\\
&&-\frac{1}{\lambda} - \frac{1}{\lambda^2} +o\left(\frac{1}{t}\right),
\end{eqnarray*}
which results in (\ref{eq:var_edges_discussion}). Statement (\ref{eq:distr_edges_discussion}) is proved along similar lines: we apply the expansion directly to the random variable
\[\frac{X_t}{t-X_t+1}=\frac{(t+1)(\lambda+1)}{(\lambda t+\lambda+1)(1-Z_t\frac{\sqrt{\lambda t}}{\lambda t+\lambda+1})}\mathbbm{1}_{\{Z_t\le \sqrt{\lambda t}\}}-1, \]
and then use the Chernoff bounds and the CLT to obtain the result.
\end{proof}

\subsection*{A3. Proof of Lemma~\ref{lem:distribution_component_sizes_one_step}}
\begin{proof}
Assume the arrival at time $t+1$ is of type $T1$. This occurs w.p. $\frac{\lambda}{\lambda+1}$, and then a new component consisting of size one is created in $G_{t+1}$, corresponding to the first case in (\ref{lem:distribution_component_sizes_one_step}). 

Next, consider a $T2$ arrival, which occurs w.p. $\frac{p}{\lambda+1}$. We now add a node to an existing component $C_i$ w.p. $\frac{|C_i|}{|V|}$. Thus the probability that we add the new node to $C_i$ is $\frac{p}{\lambda+1}\cdot\frac{|C_i|}{|V|}$.

Last, we consider a $T3$ arrival. In this case we have two options. The new edge can either join two components, or join two nodes that are already in one component. For the first case, we derive the probability that $C_i$ and $C_j$ join as
\[\begin{aligned} 
&\mathbb{P}\left(C_i\mbox{ and }C_j\mbox{ merge}\right)=\frac{1-p}{\lambda+1}\cdot\frac{2\cdot|C_i|\cdot|C_j|}{|V|^2-|V|}. \end{aligned}\]
Then for the second case, the number of ways a T3 arrival links two nodes that are already connected in a component, say $C_i$, is $|C_i|\left(|C_i|-1\right)$. Therefore with probability $\frac{\sum_{k=1}^m|C_k|^2-|C_k|}{|V|^2-|V|}$ the component size does not change.
\end{proof}

\flushend
\end{document}